\documentclass[12pt]{article}
\usepackage{epsfig}

\usepackage{amsmath}
\usepackage{amssymb}
\usepackage{euscript}

\begin{document}

\begin{titlepage}
 
\begin{flushright}
\bf CERN-PH-TH/2007-041\\
\bf TPJU-1/2007 
\end{flushright}
\vspace{2mm}

\begin{center}
{\LARGE\bf $Z$ boson as ``the standard candle''\\
            for high-precision $W$-boson physics\vspace{2mm}\\
             at LHC} \\
\end{center}

\vspace{2mm}


\begin{center}
{\large\bf  M.~W.\ Krasny$^{a}$, 
            F.\ Fayette$^{a}$, 
            W.\ P\l{}aczek$^{b,a}$
             {\rm and} A. Si\'odmok$^{b,c}$   }

\vspace{4mm}
{\em $^a$LPNHE, Pierre and Marie Curie Universit\'es Paris VI et Paris VII,\\ 
         Tour 33, RdC, 4, pl. Jussieu, 75005 Paris, France}\\  \vspace{2mm}
{\em $^b$Marian Smoluchowski Institute of Physics, Jagiellonian University,\\
         ul.\ Reymonta 4, 30-059 Cracow, Poland}\\ \vspace{2mm}
{\em $^c$CERN, Theory Division,\\
         CH-1211 Geneva 23, Switzerland}\\ 
\end{center}

\vspace{3mm}
\begin{abstract}

In this paper we propose a strategy for measuring the inclusive $W$-boson 
production
processes at LHC. This strategy exploits  simultaneously the
 unique 
flexibility  of the LHC collider in running  variable  
beam particle species at 
variable beam energies, and the configuration flexibility 
 of the LHC detectors. 
We propose their concrete settings for  a precision measurement   
of the 
Standard Model
parameters. 
These dedicated settings  optimise the use of  the $Z$ boson
and Drell--Yan-pair production processes as 
{\em ``the standard reference candles''}. 
The presented strategy allows to factorise  and to directly measure those of 
the QCD
effects which affect differently  the $W$ and $Z$ production processes.  
It reduces to a level of ${\cal O}(10^{-4})$ the impact of uncertainties in  
the  partonic distribution functions (PDFs)  and in the transverse momentum of 
the quarks on the measurement precision.  
Last but not the least, it  reduces  by a factor of $10$ an impact 
of systematic
measurement errors, such as the energy scale and the measurement resolution,  
on the $W$-boson production observables.
\end{abstract}

\vspace{0mm}
\begin{center}
{\it To be submitted to the European Physical Journal C}
\end{center}
 
\vspace{2mm}
\begin{flushleft}
{\bf CERN-PH-TH/2007-041,~ TPJU-1/2007\\
     February~2007}
\end{flushleft}  

\vspace{2mm}
\footnoterule
\noindent
{\footnotesize
$^{\star}$The work is partly supported by the program of co-operation 
between the IN2P3 and Polish Laboratories No.\ 05-116, 
and 
by the EU Marie Curie Research Training Network grant 
under the contract No.\ MRTN-CT-2006-035505.
}

\end{titlepage}

\section{Introduction}
\label{introduction}

The LHC collider will soon become  the unique $W$ and $Z$-boson production 
factory.
For the first time the collected number of the $W$-boson production events 
is expected to  be
limited by the available event storage band-width rather than by their 
production rate. About 300 million $W$ and 20 million $Z$ events are expected 
to be collected over one year of  LHC operation at the nominal luminosity.
High-precision  studies  of their static properties, their propagation 
in vacuum and in hadronic matter, their hard interactions with 
matter and with the radiation quanta  are expected to  provide the  decisive 
experimental insight into the mechanism governing   
the electroweak symmetry breaking.

One of the major challenges for the above  exploratory program is to design 
measurement  strategies which are both robust and assure
the highest--achievable precision in  controlling  the detection 
and reconstruction systematic biases. 

The robustness 
of a strategy can be quantified  in terms of its sensitivity 
to the modelling ambiguities of the effective LHC wide-band-partonic beams 
-- in particular to the modelling of their strong interaction ``noise''. 
A  robust strategy should, in our view,  be insensitive to 
the modelling aspects of their  colour and flavour dependent  properties.
In such a  strategy the effects
of  strong interactions of partonic beams, whether or not they are controlled 
by  perturbative QCD,
and the effects of their flavour
composition  must be  factorised
from the effects of  propagation, interactions and decays of electroweak bosons
and measured  rather than modelled. 
The above  factorisation is, in our view,  
particularly important  for  the first round of measurements at 
LHC. It assures that even small novel electroweak effects 
are not erroneously absorbed in the modelling ambiguities of the flavour 
composition and  strong 
interactions of partonic beams.

The ultimate, high systematic precision  measurements 
must not only use well-understood, precisely calibrated 
and aligned  detector but, in addition, the 
measurement strategy  must be carefully planned, such that the 
remaining uncertainties
of the detector response have the smallest possible  impact on the
measurement  precision.  Special care must be taken to 
avoid systematic biases driven by the  correlation 
between the systematic effects
of the detector response and those related to incomplete (approximate) 
physics modelling. Last but not the least, 
the systematic precision of the measurements 
must be validated   by the ``systematic-control" measurements
using dedicated modes of the LHC and detector operation.

This paper presents an attempt to define such a robust strategy 
for high-precision
studies of the inclusive $W$-boson production, propagation and decay processes.
 The proposed strategy 
uses  the $Z$-boson and the lepton-pair  production processes as 
{\em ``the standard reference candles''}, 
and exploits the capacities  of the LHC collider 
to run variable beam particle species at 
variable collision energies as well as 
its detectors to run in dedicated operation modes. 
The observables introduced in this paper exploit such a flexibility  
to   minimise the impact of the relative modelling 
and measurement uncertainties of the $W$ and $Z$-boson 
production and decay processes
on the measurement accuracy.

The paper is organised as follows.
In Section \ref{production}  the sources of asymmetries in measuring  
the $W$ and $Z$-boson production and decay processes  at hadronic colliders 
are traced. 
In Section \ref{measurement} the elements of the  strategy attempting  
to reduce the influence of the above  asymmetries on the measurement precision 
are presented.  
In Section \ref{monte}  the Monte Carlo tools used in the evaluation 
of the merits of the presented strategy are presented, 
while in Section \ref{detector}  
the detector model and the measurement simulation method are  discussed.
Two numerical examples of  the gains in measurement precision of the 
dedicated observables are presented in Section \ref{reduction}.
Finally, in Section \ref{factorisation}    
the method of factorisation of those of the QCD effects
which influence differently the  
$W$ and $Z$-boson production processes  is proposed, 
and the method of measurement of these effects is discussed.

\section{$W$ and $Z$ bosons at hadronic colliders}
\label{production}

\subsection{Introductory remarks}

If the partonic structure of the LHC beams was fully controlled by QCD,  
the  $Z$ and $W$-boson production rates would be predicted with
the precision limited only by the accuracy of the Standard Model (SM) 
parameters determining their masses, 
widths and couplings to leptons and quarks.
The $Z$-boson mass and  its   width
have been measured at LEP and SLC with the accuracy  of, respectively,  
${\cal O}(10^{-5})$ and ${\cal O}(10^{-3})$ \cite{PDG:2006}.
Such a precision is beyond the reach of hadronic colliders.
The $W$-boson mass and 
its decay width have been measured  with sizably inferior precision
with respect to the $Z$-bosons ones  -- a factor
of $15$ worse for the mass and a factor of $20$ worse for the decay 
width~\cite{PDG:2006}.
Any improvement in their measurement accuracy, if matched with the improvement 
on the top mass precision,   could  provide  a stringent indirect 
test of the Higgs mechanism of regularising the high-energy behaviour of 
the SM amplitudes which is 
complementary  to the direct searches of Higgs particle(s).

Several scenarios of precision measurement of the Standard Model
parameters at LHC  have been elaborated \cite{ATLAS-TDR,CMS-TDR}. 
The specificity of 
the strategy presented in this paper is that 
it introduces the same measurement procedure for  the $W$ and
$Z$-boson production processes.
Our basic goal is to minimise the extrapolation ambiguities
from the $Z$-boson production processes to the $W$-boson ones 
in order to optimally use the former as
{\em ``the high-precision standard candle''}.
The target for  such a method is to achieve a comparable precision for the 
measurements  of the $W$ and the $Z$-boson production processes.
The starting  point  is  to 
understand the asymmetries 
in the $Z$ and $W$-boson production and decay mechanisms, 
and in their detection and reconstruction methods.
The effects resulted from such asymmetries 
are  grouped in this paper in  three categories:
\begin{itemize}

\item
the physics effects,

\item
the measurement effects,

\item 
the event selection effects.

\end{itemize}

\subsection{Physics effects}

The leading process of vector  bosons production at hadron colliders is 
the Drell--Yan process of quark--antiquark annihilation. For this process, as 
well as for the sub-leading process of the $W$ and $Z$-boson bremsstrahlung by 
the decelerated
quarks, the weak-isospin  composition  of the beam particles plays 
an important role 
in creating  the $W$ and $Z$-boson production asymmetries. 
The net excess of the $u$-valence quarks  
over the $d$-valence quarks in the proton beams
is reflected mainly in the asymmetries of the rapidity distributions of the  
$W$ and $Z$-bosons. 
These asymmetries can only be partially reduced by relating  the rapidity 
distribution for  $Z$-bosons to  the  sum of distributions for the $W ^{+}$ 
and $W ^{-}$ bosons.
The remaining asymmetry is driven by the weak-isospin asymmetry
of the sea quarks which is poorly known.  
The flavour  dependent PDFs are not  
constrained by  the data in the small-$x$ region which is relevant 
for the bulk of 
the $W$ and $Z$-boson production processes at the LHC energies. They rely 
not only on the low-energy  measurements \cite{NA51:1994} but, 
more importantly,  
on the assumed $x$-dependent form of the extrapolation of
these measurement to the small-$x$ region~\cite{PDFLIB:2000}.

The difference of the masses of the $W$ and $Z$-bosons gives rise to the 
following three important
effects.  Firstly, for a given vector-boson rapidity,
different parton $x$-regions are probed for the $Z$ and  $W$ bosons. 
Any uncertainty  
in the $x$-dependence of the PDFs derived
from the QCD analysis of the deep-inelastic lepton  scattering (DIS) data is 
reflected in the  uncertainties 
of the relative rapidity distributions of  the $Z$ and $W$-bosons.
Secondly,  the resolution scale of partonic distributions is different for the 
$Z$ and $W$-boson production processes. This effect could, in principle,  
be controlled by  perturbative 
QCD,  if the partonic distributions were measured at a fixed resolution scale
with high  precision,  and if the QCD coupling constant would be known to
a very high accuracy. Since neither of the two above conditions is satisfied, 
the corresponding uncertainty  must  be taken into account for high-precision 
measurements.
Another  important strong interaction effect 
which gives rise to the $W$ and $Z$-boson production asymmetry is driven by  
the differences  in the effective transverse momentum, 
$k_T$, and the off-shellness  of partons taking part in the 
Drell--Yan process. This asymmetry is difficult to predict   because
of the interplay of the leading- and higher-twist perturbative effects  \cite{Nadolsky},
and of the non-perturbative  effects, both  determining the effective   
centre-of-collision-energy-dependent 
partonic emittance. The emittance of the LHC partonic beams could be 
measured in dedicated LHC runs using hybrid, partially stripped 
ion beams, as  proposed in  \cite{ELECTRON}. 

The differences between the masses of the down-type  and the up-type quarks,
amplified by  the CKM mixing of the down-quark flavours
contributes as well  to the  asymmetry in the production rates and 
rapidity distributions  of the $Z$ and $W$-bosons. 
This asymmetry, often neglected, 
must  be taken for consideration for high precision 
relative measurements in the $Z$ and $W$-boson sectors.

The Charged Current (CC) coupling of quarks to $W$-bosons is of 
the $V-A$ type while 
their  coupling to  $Z$ bosons is a coherent mixture of the $V-A$ and $V+A$ 
couplings. This difference is reflected in the asymmetries in the angular 
distributions of leptons originating from the decays 
of the $W$ and $Z$ bosons. In addition, at the LHC energies,  
the above asymmetries could be amplified by the 
asymmetry in production and propagation of the longitudinally polarised $W$ 
and $Z$-bosons.

The radiative corrections affect differently the $W$ and $Z$-boson
production and decay amplitudes.
While the effects of the QCD radiative corrections are  driven mainly  by the  
mass difference of the $W$ and $Z$-bosons, the effects of the electroweak 
radiative corrections lead to several, more subtle, effects. First of all,  
the virtual 
electroweak corrections  affect differently the $W$ and $Z$-boson absolute
 production
rates. In addition,  the radiation of photons affects differently 
the $W$ and $Z$-boson 
propagation and decay. This is mainly  driven by the  
differences in the interference pattern: 
(a) of the amplitudes for the photon emission from 
each of the charged leptons in $Z$-boson decays; 
(b) of the amplitudes for the photon emission 
from the charged lepton and the $W$ boson.

\subsection{Measurement effects}

The main difference in measuring the $W$ and $Z$-boson production processes 
is obvious: the lepton momentum can be directly measured while 
the neutrino momentum  can 
be reconstructed only indirectly, by  using the reconstructed momenta of 
all the particles produced in the collision of the  beam particles.  

In the LHC  colliding-beam environment the  majority of particles produced at 
small angles, with respect to the beam-collision axis, cannot be detected. 
Therefore,  the neutrino momentum is bound to be measured 
with largely inferior precision  when compared to 
that of the charged  lepton.  While the value of its 
transverse component  can be determined  from   the 
sum of transverse momenta of the detected  
particles, the longitudinal one can be estimated only,  
up to the two-fold ambiguity, using the narrow $W$-width approximation.
Moreover, in the LHC environment characterised by 
large event pile-up probability,  the modelling of the 
neutrino momentum reconstruction biases is bound to be more difficult
than at the Tevatron. Therefore, in our view, 
the high-precision measurements at  LHC must  
be based solely on the inclusive charged lepton(s) observables.

The differences in the distributions of the  pseudorapidity $\eta_l$ and of  
the transverse momentum $p_T^l$ for      
charged leptons coming respectively from $W$ and $Z$-boson decays are driven  
predominantly 
by the differences in the vector boson masses  and in the $p_T$ distributions 
of their parents.
Leptons coming from  $Z$-boson decays  have,  in general,  larger transverse momenta.
Their   pseudorapidities are less correlated with the rapidity 
of their parents
than the ones of
charged leptons coming from $W$-boson decays. 
The  lepton momentum-dependent  measurement errors  will thus 
affect differently the $W$ and $Z$-boson samples.

\subsection{Event selection effects}

Both the $Z$ and the  $W$-boson samples can be selected
using the single  inclusive lepton Level-1 triggers.
However,  the subsequent  on-line 
and off-line selection algorithms must use  the reconstructed momenta
of both leptons in order to reject  an excessive background 
of the conventional strong interaction processes. 
The asymmetry in the reconstruction precision and in the  resolution tails 
of the neutrino 
and in the charged lepton momenta gives rise to  the uncertainty of the 
 relative acceptance corrections for the $W$ and $Z$-boson events.
In addition, while the $W$-boson  events  do not contaminate 
the $Z$-boson sample,
the reverse may happen if one of the leptons is produced outside the detector
fiducial volume and/or if it is not identified. Finally, each of the leptons
from the $Z$-boson decay can give rise to the  Level-1 trigger charged lepton 
signatures.
Therefore $Z$-boson events will be accepted with higher efficiency than 
the $W$-boson ones,
if pre-selected by the Level-1 trigger system on the basis of the single-lepton
signatures.

\section{Measurement strategy}
\label{measurement}

The optimal use of the $Z$-boson as {\em ``the standard candle''} for the 
$W$-boson production processes
must take into account  the asymmetries discussed in 
the previous section and  organise the measurements in a way
which diminishes their significance for the 
measurements of  suitably chosen observables.

In this section we present the basic elements of the measurement 
strategy. They allow  us
to minimise the relative physics modelling, measurement
and event selection uncertainties discussed
in the previous section. These elements 
could be used  selectively, all them are technically feasible at LHC, 
only some of them  at Tevatron.

The basic elements of the presented strategy are listed below. 
\begin{itemize}  

\item 
Collect data  at the  two CM-energies:  
 $\sqrt{s_1}$ and $\sqrt{s_2}= (M_Z/M_W) \times \sqrt{s_1}$.
These two settings  allow to keep the momentum fractions of the partons 
producing the
$Z$ and $W$-bosons equal if the $W$-boson sample is collected at the 
CM-energy $\sqrt{s_1}$ and the $Z$-boson sample at the CM-energy $\sqrt{s_2}$. 

\item
Run  light isoscalar beams at LHC (for example the deuterium, 
 helium or oxygen beams)
to restore the weak-isospin  democracy of the 
beam particles, both in the valence and in the sea sector.

\item
Rescale the solenoid current while running at the two 
CM-energies $\sqrt{s_1}$ and $\sqrt{s_2}$ by a  factor $i_2/i_1 = M_Z/M_W$
to equalise (up to the effects of the QCD radiative corrections) the 
distribution of the curvature radius $\rho_l$ for charged leptons 
originating from the decays of the $Z$ and $W$-bosons.

\item 
Collect a fraction of data with no magnetic field to control the asymmetries 
in the trigger efficiencies and in the measured
track parameters due to the radiation of photons (resolving the 
relative size of the interference terms in the $W$ and $Z$-boson decays). 

\item
Perform both the lepton charge aware, and the lepton charge blind analysis  
to  mimic  the $V-A$ and $V+A$  mixing of the  $Z$-bosons to the final-state 
leptons using the $W^+$ and $W^-$ data.

\item
Use centrally produced $W$-bosons to control the asymmetries in the 
angular distribution of positively and negatively charged leptons in  
the $W^+$ and $W^-$ decays.

\item 
Apply the dedicated triggering and the data selection scheme to 
minimise the uncertainty  in the relative efficiency and the 
acceptance corrections
 for the $Z$ and $W$-boson samples of  events. This scheme consists of 
using the inclusive charged-lepton Level-1 trigger followed by 
the $Z/W$-symmetric cut  in the reconstructed
lepton-track curvature  $ \rho_l$ in the high-level trigger 
and in the off-line  event selection phases. 
The high-level trigger selection criteria for the second lepton must assure
democracy for the $W$ and $Z$-boson samples. This is achieved 
by searching  for a second, same flavour but opposite charge lepton in 
the selected  bunch crossings. If such a track is found  
to point  to the same vertex
it is removed from the charged track sample 
and  the event is flagged as the $Z$-boson event. 
The missing transverse momentum estimate   
is then based on the remaining vertex-pointing charged tracks
in a way which is identical for the $W$ and the $Z$-boson samples.

\item 

Use the  dedicated procedure to measure  
those of the QCD effects that are different for the $W$ and $Z$-bosons.
This procedure will be discussed in details in Section~\ref{factorisation}.

\end{itemize}

Several of the above elements 
require some flexibility in the machine and in the detector operation 
modes. We are aware that the proposed modes 
must  not disturb the canonical LHC research program 
requiring the highest collision energy and stable detector and TDAQ 
settings. Running flexible operation modes may simply  be postponed  
to the mature phase of the LHC operation when the major quest will 
be the precision of the dedicated measurements rather then the exploration 
of the highest achievable energy and luminosity scales.
The HERA example shows clearly that after reaching the 
luminosity increase plateau new operation modes must be tried to 
preserve the quality of the collider experimental program.

Taking data with the dedicated  triggering  scheme 
requires the dedicated preparatory effort but it 
is technically  straightforward. 
Reducing the solenoid current,  even if technically feasible,  requires
the substantial  dedicated effort in understanding the performance of the 
trackers in the new magnetic field environment. This effort  may, 
however, turn out to be very useful for  better understanding 
of the systematic errors of the LHC measurements.   
Similarly, running the LHC beams
for a small fraction of time at the $10\%$ lower energy is feasible.
It is important to note,  that a small drop of the delivered luminosity 
for these runs may be 
compensated by larger band width for the recorded  $W$ production events.
On the other hand, running one of the proposed  
low-$Z$ (the charge number) isoscalar beams in the LHC collider, 
even if technically feasible,
is not foreseen in the present LHC plans. This paper can thus be
considered as one of the  attempts to build the case for running the 
isoscalar light-ion beams in the advanced phase of the LHC collider operation. 
Let us stress that if the LHC collider can deliver the luminosity 
which scales as $L_{AA}= L_{pp}/A^2$ for the light isoscalar
ion beams,  then  the event rates containing the high-$p_T$ signatures will 
be similar
for the proton  and for the light-ion collisions. This condition is met 
e.g. for the eRHIC project
at BNL~\cite{eRHIC}. It is important to note  
that in the experimental environment of the LHC collider, characterised by  
multiple proton--proton collisions taking place in the same bunch crossing,  
light iscoscalar ions do not bring any additional complication due to   
the presence of several hard-process spectator nucleons in the beam particles.
They merely  replace  the distance  at which the multiple 
soft interactions take place from the micrometer scale to the femtometer 
scale. As a consequence, they 
change only the dispersion in the distribution of the nucleon--nucleon 
collision multiplicity per bunch crossing. 
This important feature could allow us 
to reopen,  at high luminosity,  the physics  program of the 
soft and diffractive collisions while preserving the high 
rate of hard partonic processes.

\section{Monte Carlo tools} 
\label{monte}

 The optimal, for the measurement strategy advocated in this paper,  
 $Z$-boson and $W$-boson
 Monte Carlo generators, should be  
based on  the same framework and numerical methods, identical SM parameter
representation, and the same modelling   of the QCD processes.
Such twin generators are  being presently developed
by two of us (WP and AS) \cite{WINHAC:2003,ZINHAC:MC}.

In  the present study we use the Monte Carlo event generator 
{\sf WINHAC}~\cite{WINHAC:2003}, version 1.22~\cite{WINHAC:MC} 
for the studies of the $W$ and $Z$ production and  decays.

At its present  development stage, {\sf WINHAC} 
contains only the leading-order process of creation of $W$-bosons 
but includes already the electroweak (EW) 
radiative corrections in leptonic $W$ decays.
The collinear configurations of initial quarks are generated
from the PDFs where the perturbative QCD effects are included through 
appropriate scaling violation.
Non-collinear configurations of initial quarks are generated 
at present using the   {\sf PYTHIA} LO-type parton showers~\cite{PYTHIA:2006}.
In the future they will be generated by the
dedicated constrained initial-state parton
shower algorithms, matched with the NLO contributions 
to the hard process -- such an extension is presently under development.
These aspects are  discussed in more details 
in Refs.~\cite{Golec-Biernat:2006xw,IFJPAN-V-04-06,Jadach:2005bf}.

A large array of PDF parametrisations 
is provided through the {\sf LHAPDF} package~\cite{LHAPDF}.
This standard package incorporates the nuclear PDFs
on top of those for the proton beams.
The nuclear PDFs include the nuclear shadowing effects parametrised by the EKS 
group~\cite{EKS:1999}.
This parametrisation is the DGLAP extrapolation of the DIS data taken with 
several nuclear targets to the scales involved in production of the $W$-bosons 
and to the nucleus type used in the presented studies.

Since,  at hadron colliders,  the $W$-bosons can be identified efficiently
only through their leptonic-decay channels, 
in {\sf WINHAC} such decays are the only ones which are considered.
The process of leptonic $W$-boson decays is described within the framework
of the Yennie--Frautschi--Suura exclusive exponentiation~\cite{yfs:1961},
where all the infrared QED effects are re-summed to the infinite order,
while the residual non-infrared EW corrections are calculated perturbatively.
In the current version of the program the latter corrections are included up 
to ${\cal O}(\alpha)$.
{\sf WINHAC} went successfully through several numerical 
tests~\cite{WINHAC:2003},
and was also compared with the independent Monte Carlo
program {\sf HORACE}~\cite{WINHAC-HORACE:2004}. 

In the current version of {\sf WINHAC} we have included the Born-level
neutral-current ($Z+\gamma$) Drell--Yan process. 
For the non-collinear
configurations it uses the same interface to the  {\sf PYTHIA} LO-type 
parton showers \cite{PYTHIA:2006} as in the case of the $W$-boson production. 
The QED and the EW corrections are not included in the present version of the 
program.
A dedicated event generator for the $Z$-boson production and decay, called 
{\sf ZINHAC}, is in the process of development \cite {ZINHAC:MC}.

The important  merit of the {\sf WINHAC} event generator, for  
the measurement strategy presented in this paper, 
is that the $W$ and $Z$-boson production and decay
processes are described using the  spin amplitude formalism. 
The spin amplitudes  are calculated
separately for the weak bosons  production and for their  decays.
They correspond to all possible spin configurations of the
intermediate $W$ and $Z$-bosons and the initial and final-state fermions. 
The matrix element for the charged-current Drell--Yan process is
obtained by coherently summing the production and decay amplitudes over the 
intermediate $W$ and $Z$-boson spin states. 
The amplitudes are evaluated numerically 
for given particles four-momenta and polarisations. They can be
calculated in any Lorentz frame in which the corresponding particles
four-momenta are defined, for more details see Ref.~\cite{WINHAC:2003}.
The advantage of using the spin amplitudes is that one can control the 
spin states, in particular the production of longitudinally and
transversely polarised $W$ and $Z$-bosons.

One of the novelties  of the measurement strategy  discussed  in this  paper 
is the factorisation of the electroweak
and the strong interaction effects.
For the precision measurements of the electroweak parameters the latter 
are proposed to be determined  using the dedicated procedure
based solely on the data. Their
partial inclusion in the present studies (reduced
only to perturbative ones, and implemented in the approximate LO form) 
is merely to indicate the size of the corrections and to test the 
proposed factorisation procedure. 
On the other hand all the state-of-the-art electroweak effects which 
can only be partly controlled experimentally, 
must eventually be implemented in 
the {\sf WINHAC} and {\sf ZINHAC} generators
for the high-precision studies of the high-statistics LHC data.

\section{Detector model and Monte Carlo studies }
\label{detector}

At LHC, the high-precision measurements of 
the $W$-boson production observables  will be based on the  samples 
of at least $10^8$ recorded $W$-boson production  events. 
Full detector simulation of comparable sample of Monte Carlo 
events is both unrealistic and unnecessary. The precision measurements 
will be bound to  use the correction factors (efficiencies, acceptances 
etc.) derived directly from the data and/or 
in form of the parametrised response of the detector. 

In the studies presented in this paper we use the average response functions of
the ATLAS  tracker to charged particles as specified in Ref.~\cite{ATLAS-TDR}.
We restrict our studies to particles produced in the pseudorapidity 
range $-2.5 \leq \eta _l \leq 2.5$. In the future  the response functions
of the LHC detectors  will be determined
{\em in situ} 
from the measurements of the decay products of the known narrow resonances.
The ATLAS detector response functions are used here merely for the initial
 estimate of the size of the systematic measurement effects.

We have generated, using  the dedicated processor farms,
several large samples ($10^8$ events)
of the $Z$ and $W$-boson production events using the  {\sf WINHAC} generator.  
The generated $Z$ and $W$-boson  samples  have been processed
using either the average or the  ``biased''  detector-response functions.
The studied biases included 
shifts in the scale  of the reconstructed momenta and in the  
detector resolution.
The $Z$-boson production events have been  generated at 
$\sqrt{s_n} = 14\,$TeV (proton beams) and 
$\sqrt{s_2}= 7\,$TeV/nucleon  (isoscalar ion beams). The $W$ production events
have been generated both at the CM-energy of $\sqrt{s_n}$
(proton beams) and at the energy 
 $\sqrt{s_1} =  (M_W/M_Z) \sqrt{s_2}  $ (isoscalar ion beams). 
For the samples generated   at the CM energies of $\sqrt{s_n}$ 
and $\sqrt{s_2}$ the solenoid-coil current 
$i(s)$ has been set to the nominal 
value $i_n$, corresponding to the assumed tracker response 
functions \cite{ATLAS-TDR}. 
For the samples generated at the energy $\sqrt{s_1}$ the coil current has been 
rescaled down by a factor equal to the ratio of the $W$  and $Z$-boson masses.

The $W$ and $Z$-boson events have been selected by demanding the presence
of the charged lepton with the track curvature 
radius $\rho_l $ satisfying the following conditions:

$$
\rho_l \:\leq \: \rho_l ^{c} =  \frac{i(s)}{i_n} \; \frac{1}{p_T^{c}(s)}\,,
$$ 
where 
\begin{eqnarray*}
& & p_T^{c}(s_n) = p_T^{c}(s_2) =  20\,{\rm GeV}\,,\hspace{15mm}
p_T^{c}(s_1) = p_T^{c}(s_n) \;  \frac{M_W}{M_Z}\,, \\
& & i(s_n) = i(s_2) = i_n\,, \hspace{30mm}
i(s_1) = i_n\; \frac{M_W}{M_Z}\,, 
\end{eqnarray*}
 
and the pseudorapidity $\eta_l$ within the following range 
$$
-2.5 \leq \: \eta_l \:\leq 2.5\,.
$$
 
In the case of  the $Z$-bosons 
and  the lepton-pair event samples   
we first  randomly choose  one of the two leptons
and select an event only if this lepton  satisfies the same selection criteria
 as specified above for the $W$-boson event samples. Note  that by 
specifying the selection condition in terms of the radius of 
the track curvature rather than in terms of the transverse momentum 
and by rescaling the solenoid current and CM-energy  we achieve almost  
symmetric  selection of the $W$-boson and the  $Z$-boson events\footnote{The
remaining residual asymmetry reflects the differences in the transverse 
momentum 
of the $Z$ and $W$-bosons and in the angular distributions of the charged 
leptons.}.
 
\begin{figure}
\begin{center}
\setlength{\unitlength}{1mm}
\begin{picture}(160,150)
\put(0,0){\makebox(0,0)[lb]{
\epsfig{file=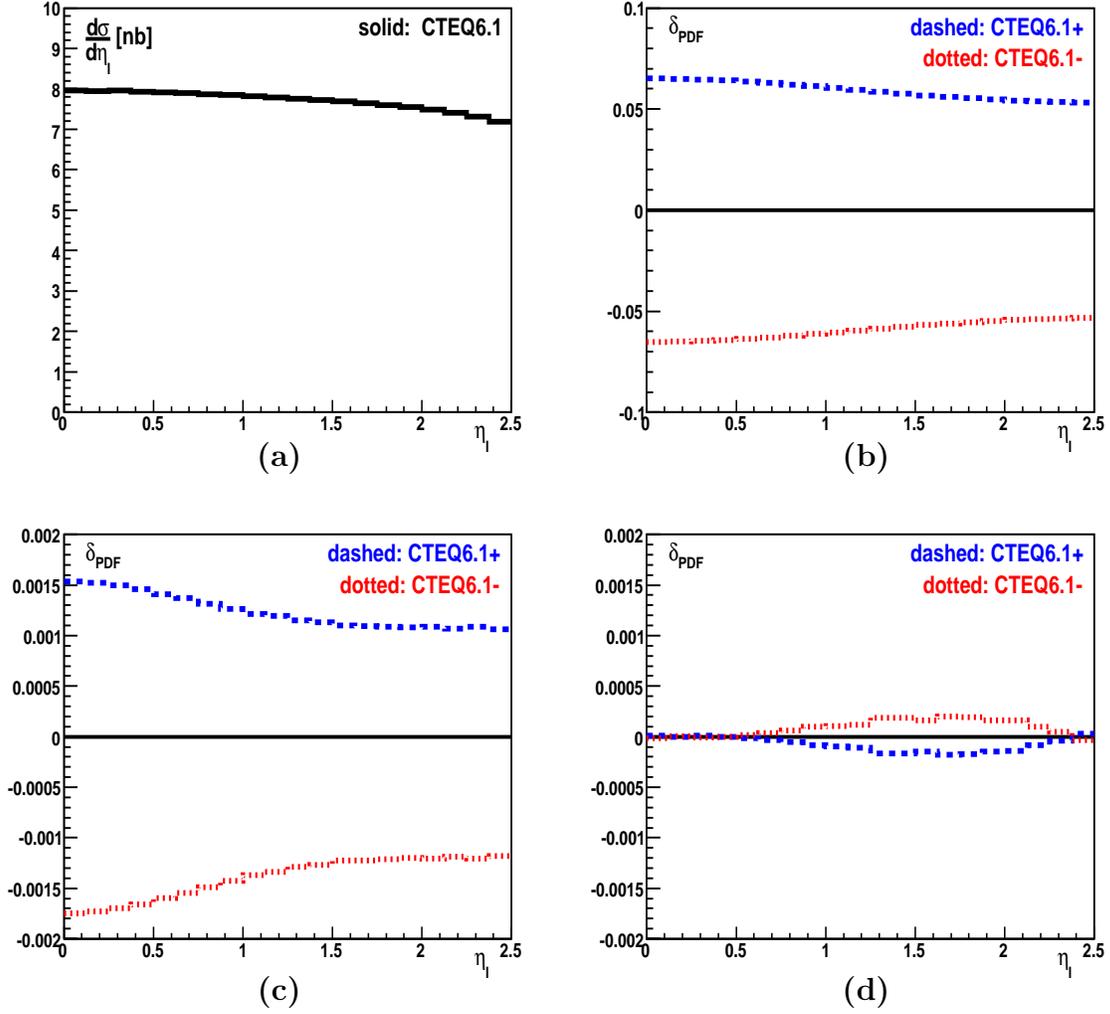,width=15.5cm,height=14cm}
}}
\put( 39,70){\makebox(0,0)[cb]{\bf (a)}}
\put(117,70){\makebox(0,0)[cb]{\bf (b)}}
\put( 39,-1){\makebox(0,0)[cb]{\bf (c)}}
\put(117,-1){\makebox(0,0)[cb]{\bf (d)}}
\end{picture}
\end{center}
\caption{\sf (a) The distribution of the lepton pseudorapidity $\eta _l$ 
for proton--proton collisions at LHC;  
(b) the systematic uncertainty  
$\delta _{PDF} = \frac{d \sigma/d \eta_{l}({\rm CTEQ6.1} \pm) 
- d \sigma/d \eta_{l}({\rm CTEQ6.1})}
{d \sigma/d \eta_{l}({\rm CTEQ6.1})}$ 
of the $\eta _l$ distribution reflecting  the PDF uncertainty;
(c) as above but for  the ratio of the $\eta _l$ distributions for
the $W$ and $Z$-boson samples;  
(d) as above but for the collision of the isoscalar beams,
the rescaled collision energy 
and the rescaled magnetic field (see the text for 
details).  }
\label{etal}
\end{figure}

\section{Reduction of systematic and modelling uncertainties -- examples}
\label{reduction}

In this section we discuss  two numerical examples of improving 
 the measurement 
precision of the dedicated $W$-boson observables. These improvements  
 can be achieved using the first three elements of the strategy  presented
in Section~\ref{measurement}.   The construction of  dedicated, 
precision-measurement, 
  observables must assure their stability  both with respect to the measurement
 biases of the lepton kinematics and with respect to the 
modelling ambiguities of the partonic beams at LHC.
Their sensitivity to the $W$-boson production, propagation and decay dynamics
must remain the same as for the classical observables.

In the present studies we do not investigate the contribution 
of the QCD background to the selected samples of the $W$ and $Z$-boson  
events. Earlier studies \cite{Haywood:2000} 
have shown that the QCD background contamination  is small  
and its  uncertainty will have a 
negligible effect on the final measurement precision.  
Moreover, we do not evaluate here the gains coming from 
those elements of the measurement 
strategy proposed in Section \ref{measurement} which aim at reducing the 
impact of the asymmetries in the relative size of the electromagnetic radiative
corrections due to real photon emissions.
The latter, of  high importance for the precision 
measurements,   will be presented 
in a separate communication,   when the implementation of
these processes in the   {\sf ZINHAC} generator is finalised.
As long as radiation processes are neglected 
the electron and the muon track reconstruction 
quality remain  the same. In 
the following, leptons will have a meaning of either electrons or  muons.

In Fig.~\ref{etal}a  we show the charged lepton 
pseudorapidity distribution for the 
$pp \rightarrow W + X$,
$W \rightarrow l \nu$ process 
at the CM energy of $\sqrt{s_n}$,  for the CTEQ6.1 partonic density 
distributions \cite{CTEQ6.1:2003}.
The contribution to the uncertainty of this distribution coming  
from the uncertainties in the partonic distribution functions (PDFs)  
and determined using the method proposed in \cite{CTEQ6:2002} is shown in 
Fig.~\ref{etal}b.
This uncertainty may be diminished to the per-mil level,
as shown in Fig.~\ref{etal}c,  
by replacing the pseudorapidity distribution  by the ratio of
the charged lepton pseudorapidity distributions  
for the $W$ and $Z$-boson production events.

\begin{figure}
\begin{center}
\setlength{\unitlength}{1mm}
\begin{picture}(160,150)
\put(0,0){\makebox(0,0)[lb]{
\epsfig{file=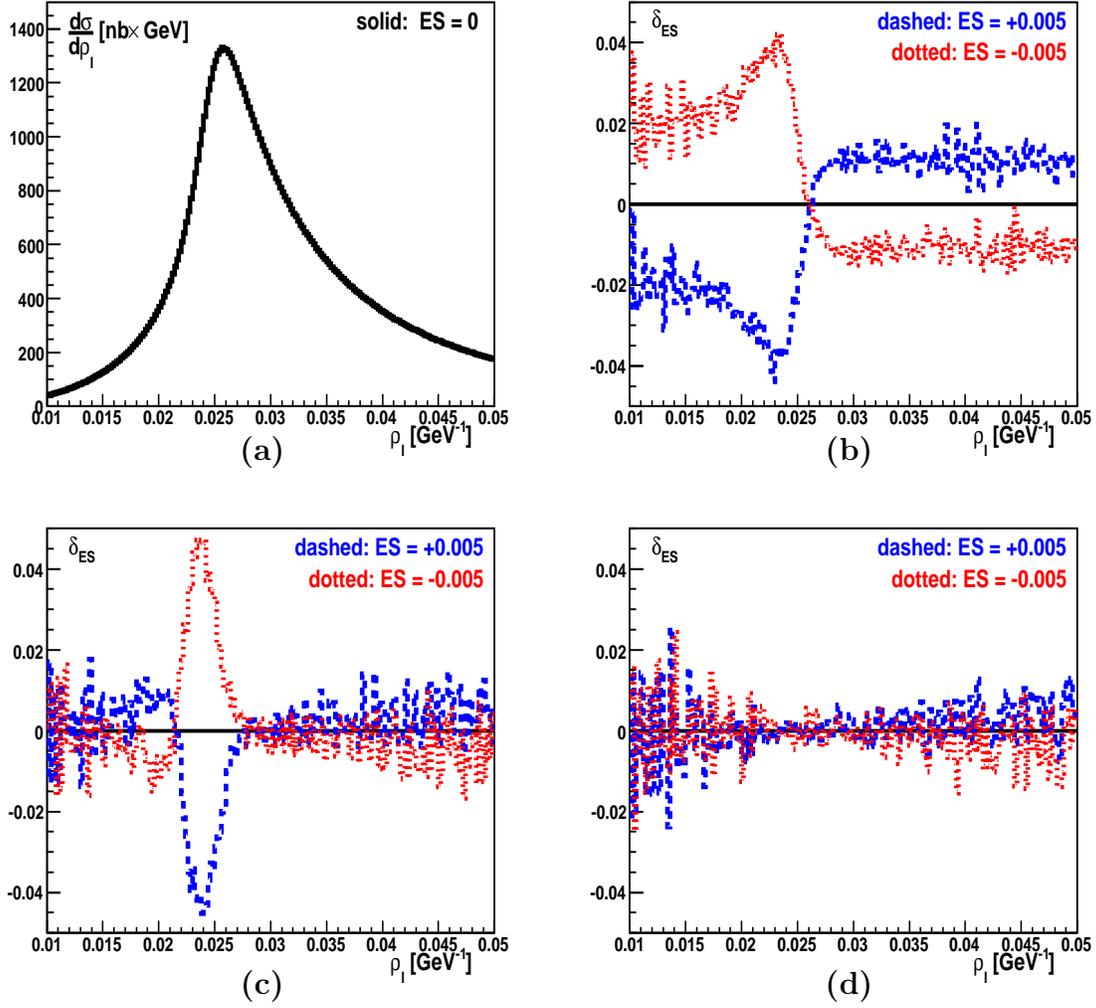,width=15.5cm,height=14cm}
}}
\put( 39,70){\makebox(0,0)[cb]{\bf (a)}}
\put(117,70){\makebox(0,0)[cb]{\bf (b)}}
\put( 39,-1){\makebox(0,0)[cb]{\bf (c)}}
\put(117,-1){\makebox(0,0)[cb]{\bf (d)}}
\end{picture}
\end{center}
\caption{\sf 
(a) The distribution of  the lepton track curvature radius $\rho_l$ 
for proton-proton collisions at LHC; 
(b) the  systematic  uncertainty
$\delta _{\rm ES} = 
\frac{d \sigma/d \rho_l({\rm ES} \pm) - d \sigma/d \rho_l({\rm ES})}
{d \sigma/d \rho_l({\rm ES})}$ of the 
$\rho _l$ distribution generated by the lepton-momentum scale uncertainty;
(c) as above but for  the ratio  of the $\rho _l$ distributions for
the $W$ and $Z$-boson samples; 
(d) as above but for the collision of the isoscalar beams,
the rescaled collision energy and the rescaled magnetic field 
(see the text for details).
}
\label{rhol}
\end{figure}

For further reduction of the impact of the uncertainty of the 
PDFs we propose to measure the 
isoscalar beam collision  observable,   defined as:
\begin{equation}
R^{iso}_{WZ} = 
\frac{d \sigma^{iso}_W (s_1, i(s_1))}{ d \sigma^{iso}_Z(s_2,i(s_2))}\,. 
\label{eq:RWZiso}
\end{equation} 
 This observable  is plotted in Fig.~\ref{etal}d as a function 
 of the lepton pseudorapidity for the deuterium beams.
 Its sensitivity to the uncertainty in  the partonic
 distribution functions is reduced to a  level below  
 $2 \times 10^{-4}$. 
            This residual uncertainty is driven predominantly by the
            CKM mixing of down-quark flavours and by
            the differences of masses
            of the down- and up-type quarks. The uncertainty due to
            the asymmetry in the shadowing effects for  the $W$ and
            $Z$-boson production is negligible owing to the smallness
            of the shadowing effects
            for light ions, their weak-isospin invariance, and their very
            mild resolution-scale dependence in the vicinity of the $M_W$
            scale.

 Note that the $W$ and $Z$-boson samples would have to be  collected
 over the distinct  beam running periods.  The quoted accuracy 
 concerns thus the shape of  the $R^{iso}_{WZ}$ ratio. The precise
 normalisation of $R^{iso}_{WZ}$, which is sensitive  
 to  the ratio of $\Gamma _W / \Gamma _Z$ requires  
 the  dedicated method of absolute cross-normalisation of the event
samples collected at the  two CM-energies. Such a scheme,
aiming at the per-mil precision,  is  being developed
\cite{KChS:2006}. Another method of getting rid  
of the uncertainty in  the absolute normalisation of $R^{iso}_{WZ}$
is presented in the next section.

The dominant factor limiting the precision of the measurement 
of the $W$-boson mass is the uncertainty in  the scale of the 
lepton transverse momentum (energy). To improve its present precision,
the lepton energy and momentum scale must  be known to 
better than 0.02\% \cite{Haywood:2000}. The measurement strategy
discussed in this paper  allows us to drastically reduce the 
influence of the scale error on the measurement of the $W$-boson mass.

In Fig.~\ref{rhol}a we present the distribution of the curvature
radius $\rho_l$ of the lepton track originated from the decays 
of the $W$-bosons
produced in the $pp$ collisions at the nominal LHC energy. The 
peak position and the shape 
of this distribution is sensitive both to the $W$-boson mass
and to the lepton momentum scale bias.  In Fig.~\ref{rhol}b
we  show the effect of the change in  the scale of the reconstructed  
lepton transverse momentum: $p_T ^{rec} = p_T ^{true}(1 \pm \varepsilon_s)$ 
for $\varepsilon_s = 0.005$. A scale uncertainty of this magnitude leads to  
errors  of up to $4\%$ of the $\rho_l $ distribution which, in turn, gives rise
to the uncertainty of the measured $W$-boson mass of about $500\,$MeV. 
This uncertainty is only slightly reduced  when normalising the distribution
for $W$-bosons to the corresponding one for $Z$-bosons,  as shown in  
Fig.~\ref{rhol}c.

The $R^{iso}_{WZ}$ observable, if measured in 
dedicated runs of isoscalar beams at the two CM-energy and solenoid-current 
settings, allows us to drastically reduce the above  uncertainty. 
This observable is  plotted in   Fig.~\ref{rhol}d 
as a function of $\rho_l$.
In the peak region sensitive to the $W$-boson mass
the impact of the  lepton momentum scale uncertainty
on the   $R^{iso}_{WZ}$ observable
is reduced by the factor of $10$ with respect to the direct measurement 
of the $\rho_l$ distribution in the nominal-energy proton-beam runs.

\section{Factorisation of QCD effects}
\label{factorisation}

The rescaling of the energy of the LHC beams 
allows us to consider  the formation  of the $Z$ and $W$-bosons  on the 
same footing --  for a given rapidity of the $Z$ and $W$-bosons the fractions
of the proton momentum carried by annihilating partons are, by 
construction, the same. 
However, this holds exactly only for collinear massless partons.

\begin{figure}
\begin{center}
\setlength{\unitlength}{1mm}
\begin{picture}(160,150)
\put(0,75){\makebox(0,0)[lb]{
\epsfig{file=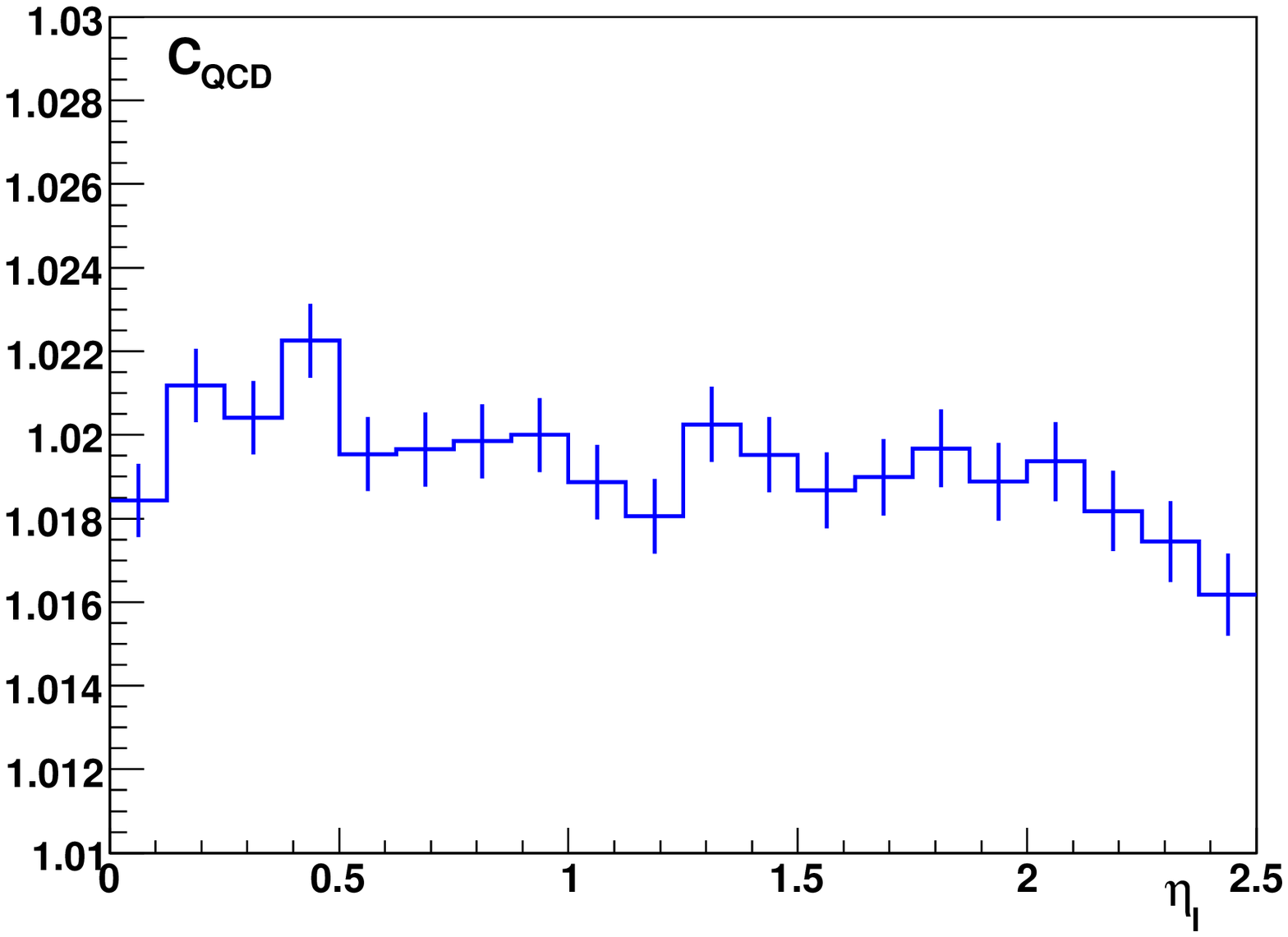, width=80mm,height=70mm}
}}

\put(0, 0){\makebox(0,0)[lb]{
\epsfig{file=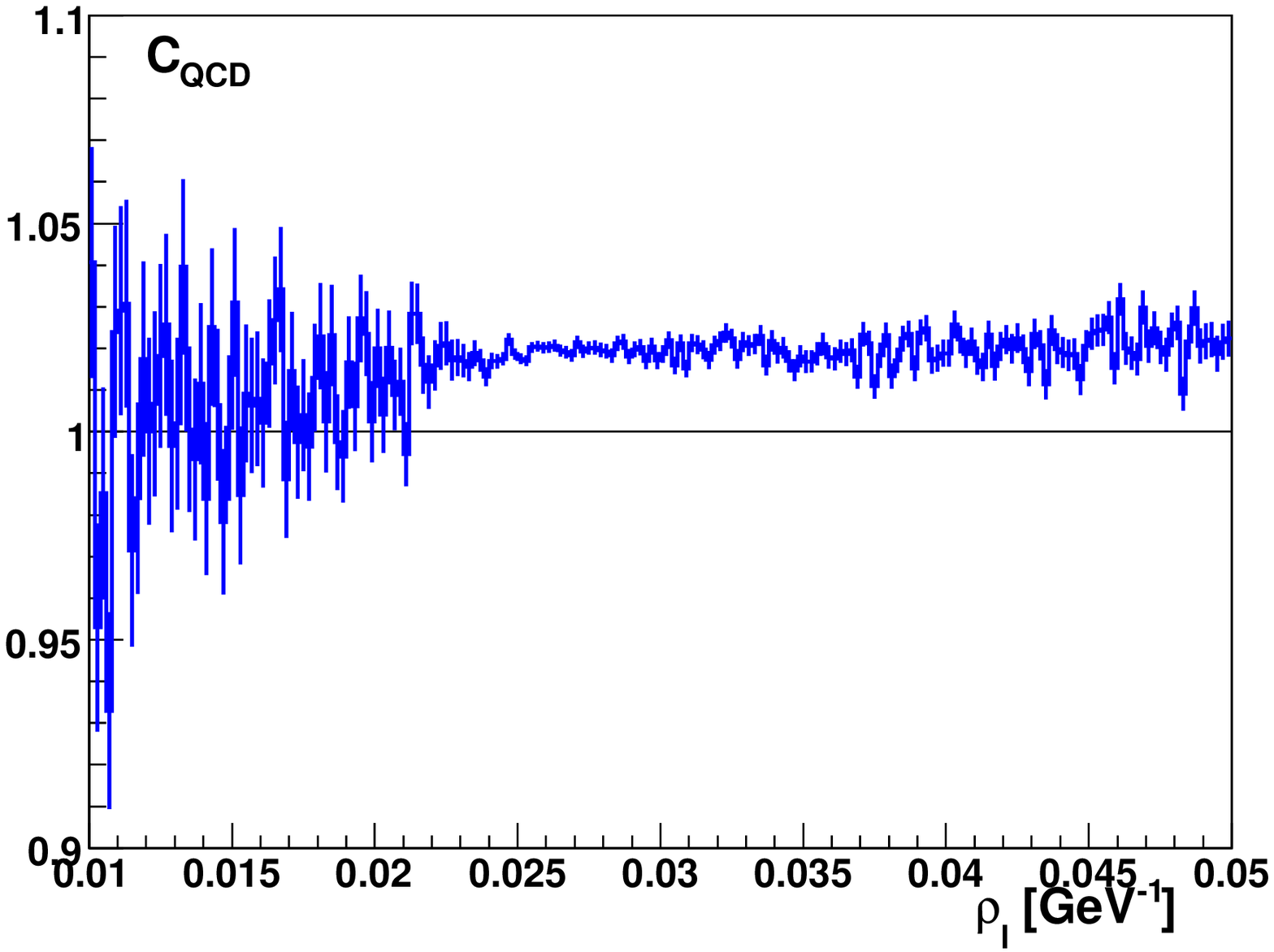, width=80mm,height=70mm}
}}

\put(75,75){\makebox(0,0)[lb]{
\epsfig{file=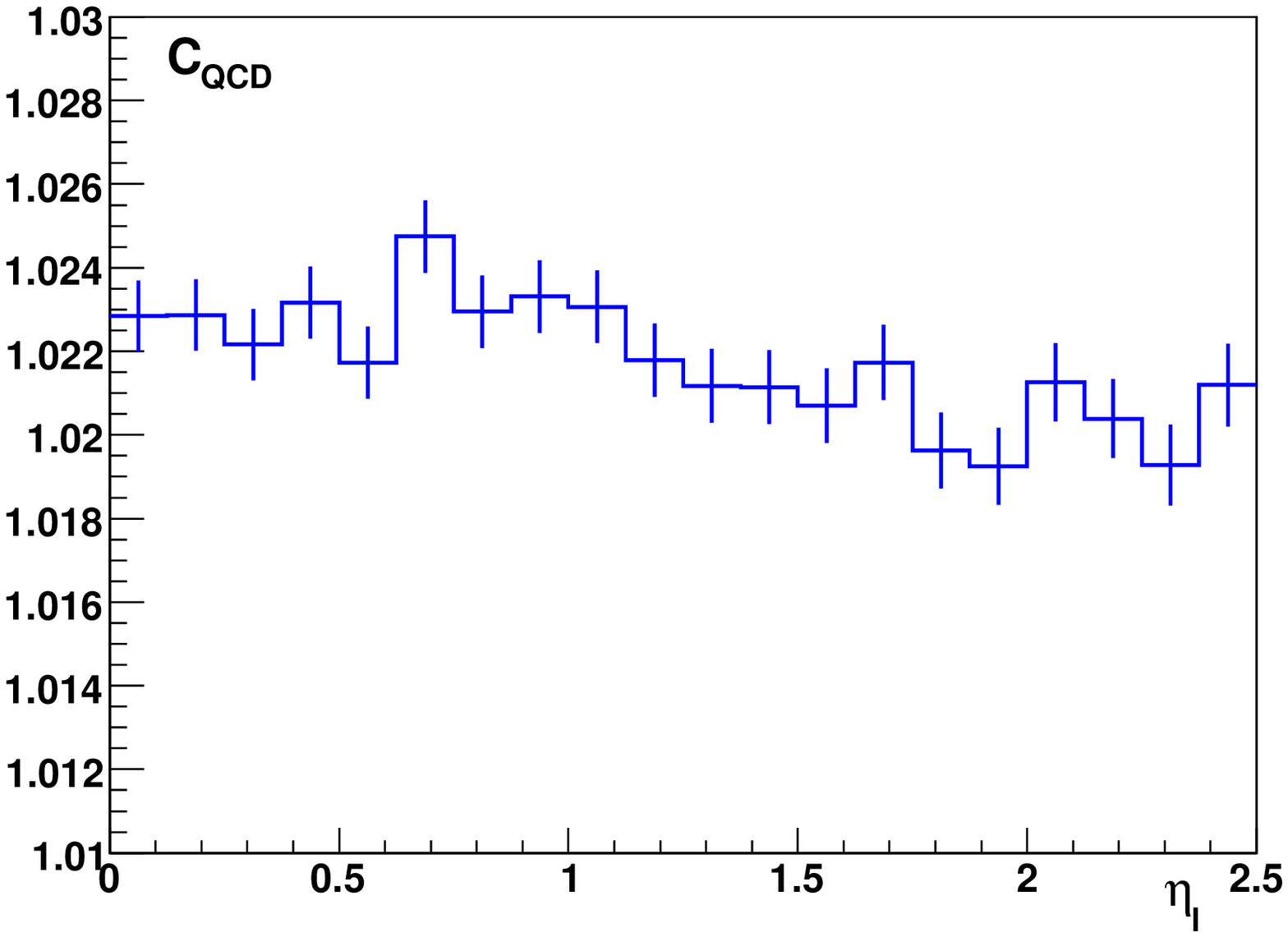, width=80mm,height=70mm}
}}

\put(75, 0){\makebox(0,0)[lb]{
\epsfig{file=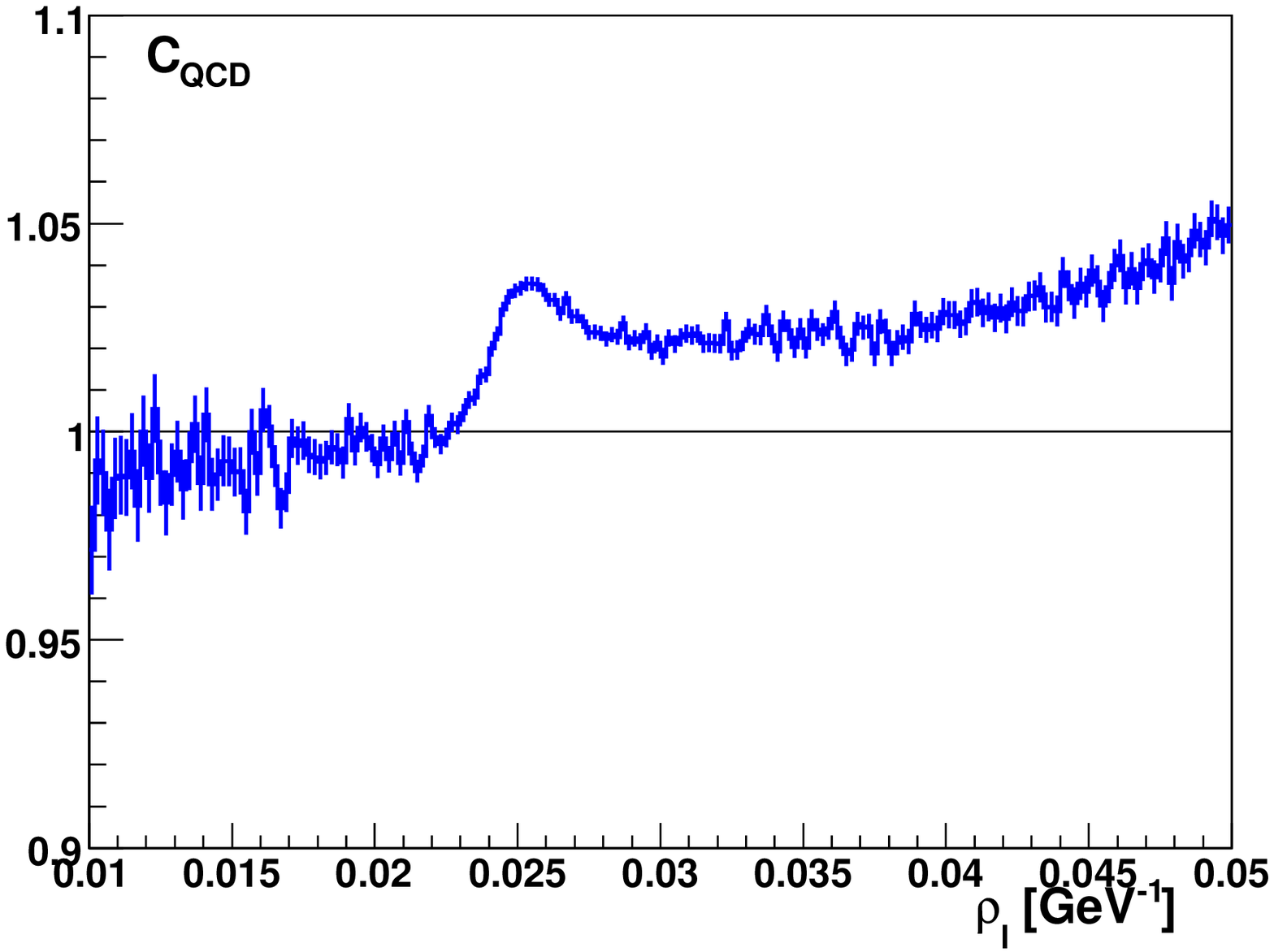, width=80mm,height=70mm}

}}
\put( 40,72){\makebox(0,0)[cb]{\bf (a)}}
\put(115,72){\makebox(0,0)[cb]{\bf (b)}}
\put( 40,-3){\makebox(0,0)[cb]{\bf (c)}}
\put(115,-3){\makebox(0,0)[cb]{\bf (d)}}

\end{picture}

\end{center}

\caption{The  $C_{QCD}(\eta_l)$  and $C_{QCD}(\rho_l)$
correction functions  (see the text for detailed  explanation) for 
collinear partons -- (a) and (c), respectively, and for {\sf PYTHIA} modelling 
of their transverse momenta -- (b) and (d), respectively. }
\label{qcd}
\end{figure}

In reality,  the   $R^{iso}_{WZ}$  observable defined  
in Section~\ref{reduction} 
is  still sensitive to several effects which must  be 
experimentally controlled for high-precision measurements.
Firstly, it is  sensitive to  the scale dependence  
of  partonic distributions, which is governed  by the  QCD coupling 
constant $\alpha _s(M_W)$ and by the scale-dependent factor $\ln(M_W^2/M_Z^2)$.
It is  also sensitive to the effective  
distributions of the transverse momenta $k_T$ and the off-shellness 
$m_{\ast}$ of the 
annihilating quarks,  which can only be partly controlled
by perturbative QCD.  
Last but not the least,  the  $R^{iso}_{WZ}$ observable is sensitive
to the relative normalisation  of the $Z$ and $W$-boson samples taken 
in separate runs.

In order to get rid of the above effects, rather than to model them,  
we propose 
to select the samples of events containing a pair of opposite charge
and same flavour leptons, and to  measure the ratio of the integrated 
lepton pair production rates  
\begin{equation}
C_{QCD} = 
\frac{\int_{M_Z- 3\Gamma _Z}^{M_Z+ 3\Gamma _Z}N^{l+l-}(s_2, i(s_2), M^{l+l-} )
\;dM^{l+l-}} 
 {\int_{M_W- 3\Gamma _W}^{M_W+ 3\Gamma _W} 
 f_{BW} (s^{l+l-}; M_W ,\Gamma _W) \; w_{EW} 
 \;  N^{l+l-}(s_1,i(s_1),M^{l+l-} ) \;dM^{l+l-}}
\label{eq:CQCD}
\end{equation}
as a function of $\rho_l$ and as a function of $\eta_l$ 
of the randomly chosen lepton. 
The rates $N^{l+l-}$ in the above formula are integrated over
the invariant mass $M^{l+l-}$ of the lepton pairs in the regions 
$( M_Z - 3\Gamma _Z, M_Z +3\Gamma _Z)$,  
and $( M_W - 3\Gamma _W, M_W +3\Gamma _W)$,
correspondingly. Each event having a reconstructed invariant mass in the 
latter region is weighted by the Breit--Wigner function 
$$
f_{BW} (s^{l+l-}; M_W ,\Gamma _W) =  \frac{1}{\pi}\;
\frac{M_W\Gamma _W}{(s^{l+l-} - M_W^2)^2 + M_W^2\Gamma _W^2}\,,
$$ 
where $s^{l+l-} = (M^{l+l-})^2$,
and by 
the QCD-independent normalisation factor $w_{EW}$. 
This factor is defined such that 
the integral of the weighted lepton invariant mass spectrum in the region  of  
 $( M_W - 3\Gamma _W, M_W +3\Gamma _W)$ is equal to the cross section of 
a $Z$-like boson having the mass and the width of the
 $W$-boson\footnote{The weighting procedure
 takes care  of the residual asymmetries
 of the angular distributions of the leptons produced in the region of 
 the $Z$-peak and in the region outside the $Z$-peak. }.
The lepton pair production events used in the determination of the 
$C_{QCD}(\eta_l)$  and $C_{QCD}(\rho_l)$
correction functions 
must be triggered and selected on the basis of the presence of 
two same-flavour, oposite-charge lepton candidates.
Each of the lepton must satisfy the kinematical selection 
criteria specified in Section~\ref{detector}.
This requirement, stronger than the corresponding one for the $Z$-boson sample 
of events discussed in 
Section~\ref{detector}, 
is necessary to reduce the background to the inclusive lepton
samples in the lepton-pair invariant mass region outside the $Z$-peak.

\begin{figure}
\begin{center}
\setlength{\unitlength}{1mm}
\begin{picture}(160,150)

\put(0,75){\makebox(0,0)[lb]{
\epsfig{file=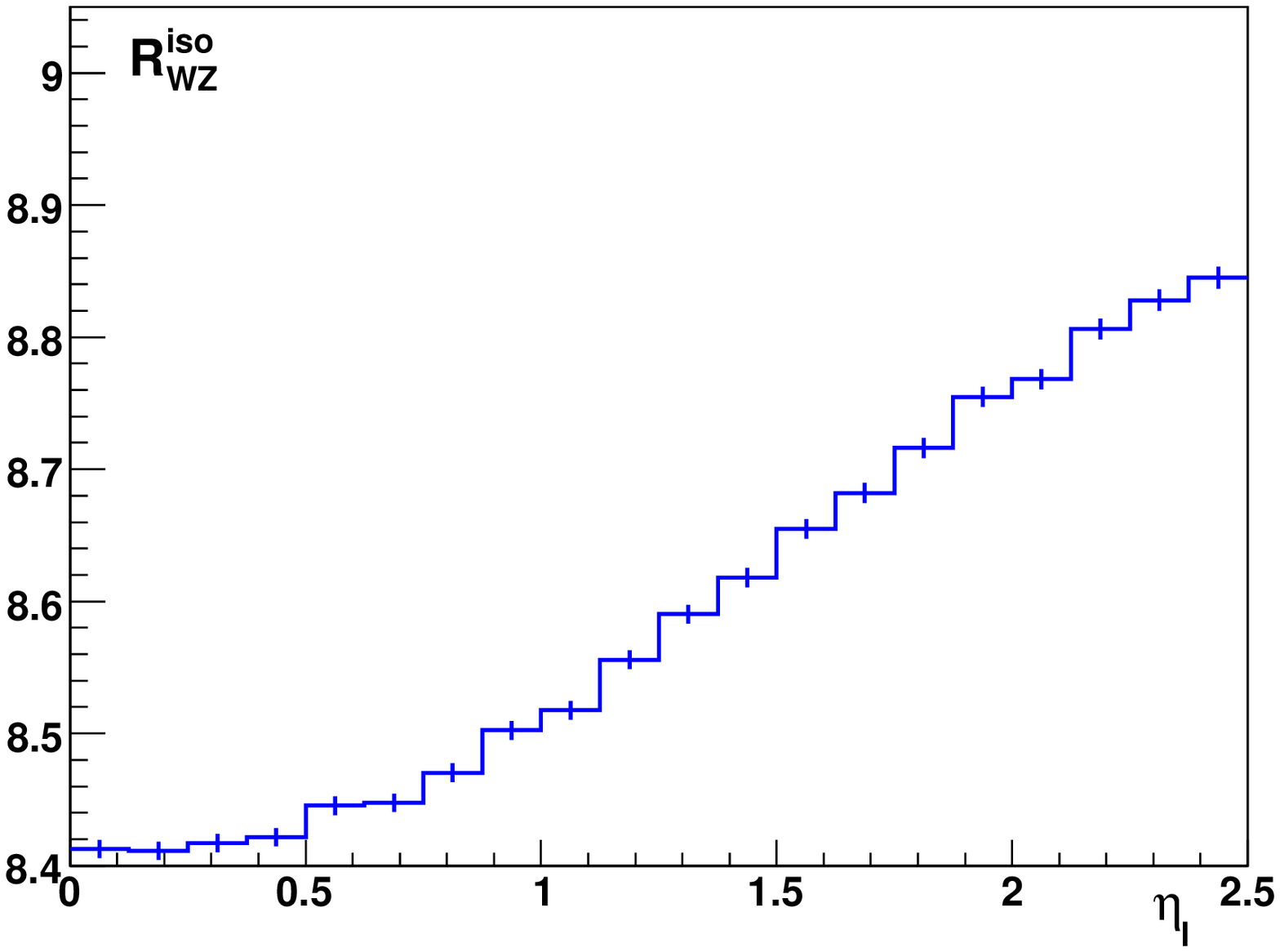, width=80mm,height=70mm}
}}

\put(0, 0){\makebox(0,0)[lb]{
\epsfig{file=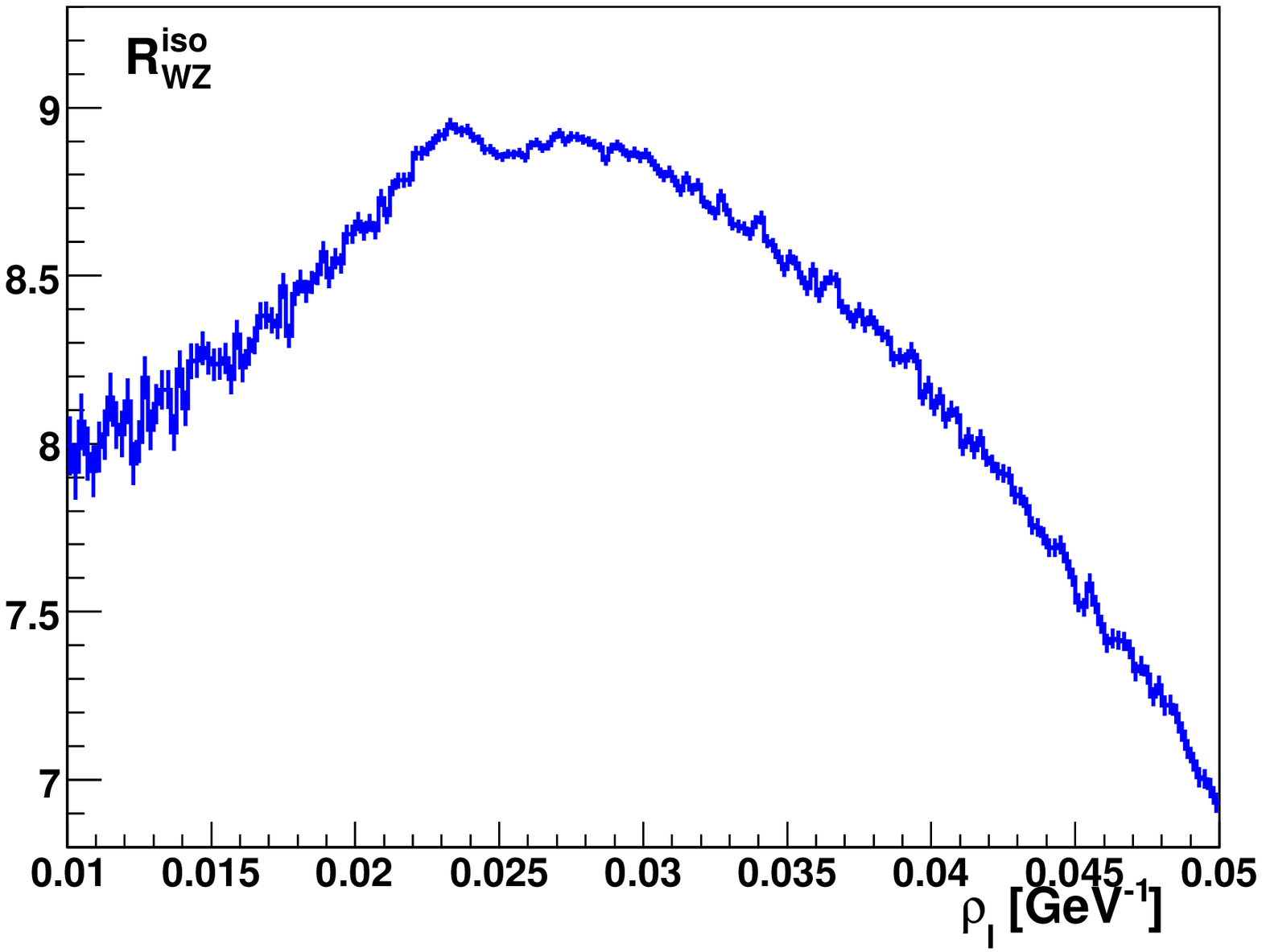, width=80mm,height=70mm}
}}

\put(75,75){\makebox(0,0)[lb]{
\epsfig{file=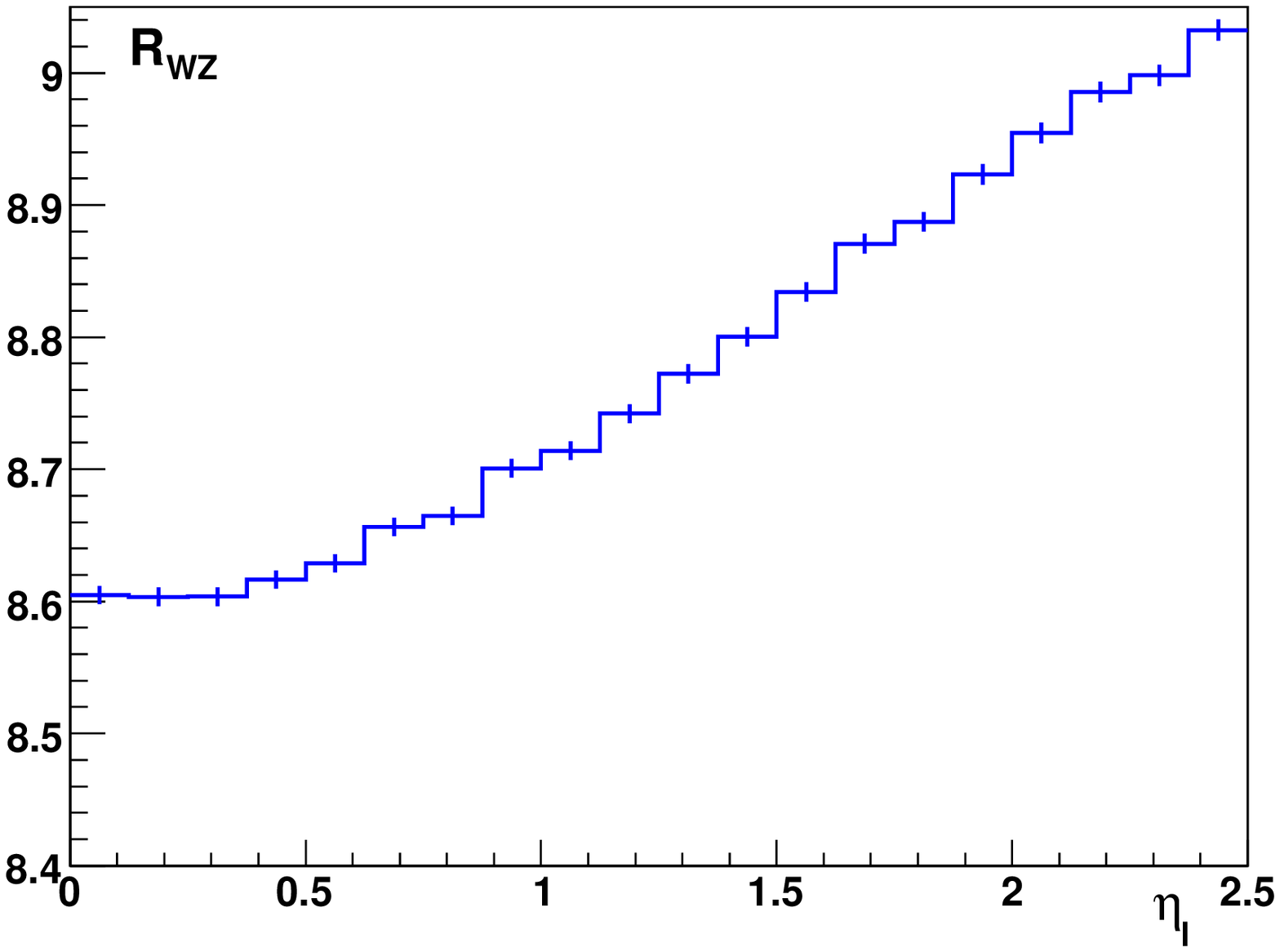, width=80mm,height=70mm}
}}

\put(75, 0){\makebox(0,0)[lb]{
\epsfig{file=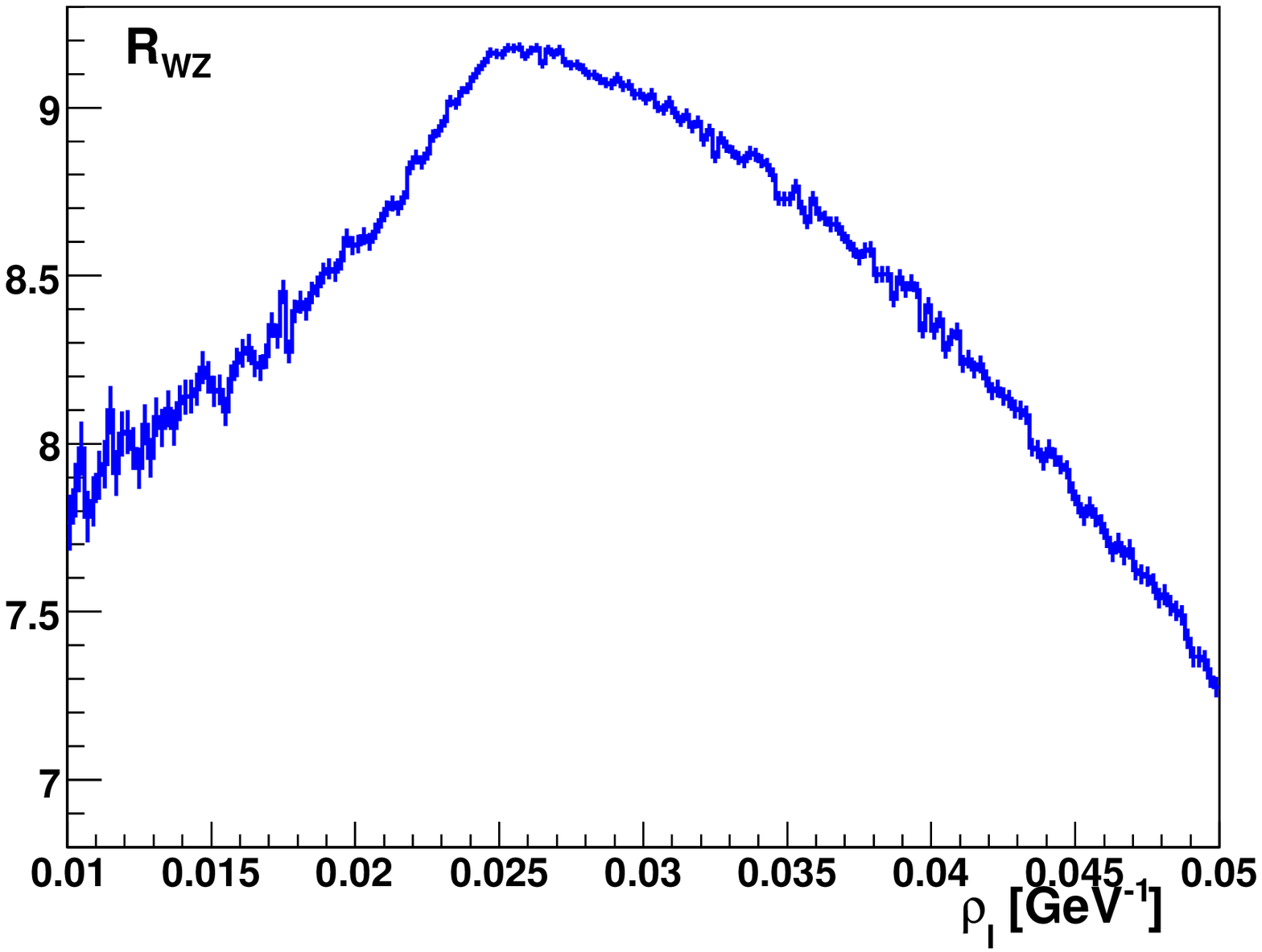, width=80mm,height=70mm}
}}
\put( 40,73){\makebox(0,0)[cb]{\bf (a)}}
\put(115,73){\makebox(0,0)[cb]{\bf (b)}}
\put( 40,-2){\makebox(0,0)[cb]{\bf (c)}}
\put(115,-2){\makebox(0,0)[cb]{\bf (d)}}

\end{picture}

\end{center}
\caption{\sf Uncorrected ratios $R^{iso}_{WZ}$ and the QCD-corrected ratios 
              $R_{WZ}$ plotted as a function of  
              $\eta_l$ -- (a) and  (b), respectively, 
              and as a function of $\rho_l$ -- (c) and  (d), respectively.
         }  
\label{qcd1}
\end{figure}

The $C_{QCD}(\eta_l)$  and $C_{QCD}(\rho_l)$ correction functions
 will be  be  determined directly from the lepton pair
 production data  for  given input  functions  
$w_{EW}$ and  $f_{BW}(s^{l+l-}; M_W ,\Gamma _W)$. 
 Since these  functions are sensitive to the SM
 parameters they should, in principle,  be calculated iteratively.
 In the first iteration step the PDG values of the parameters can be used. 
The corresponding 
 correction functions  $C_{QCD}$ would then be used to determine 
their more precise 
 values from the analysis of the QCD-effects corrected  $R^{iso}_{WZ}$
 observables. The improved values could then be used in the next iteration
 step.

 In Fig.~\ref{qcd} we present our  initial estimate of the size 
of the correction functions
 $C_{QCD}(\eta_l)$  and $C_{QCD}(\rho_l)$, 
 using the CTEQ parametrisation of the PDFs and  the corresponding value 
of  the QCD coupling constant \cite{CTEQ6.1:2003}. 
In order to understand the relative size of the 
 scaling violation effects and the effects due to  
the transverse momentum of the quarks,
 the $C_{QCD}(\eta_l)$  and $C_{QCD}(\rho_l)$ correction functions have been 
 determined first   
 assuming $k_T = 0$ and $m_{\ast} = 0 $ 
 -- Fig.~\ref{qcd}a and   Fig.~\ref{qcd}c, respectively, 
 and then  using  their {\sf PYTHIA} modelling \cite{PYTHIA:2006} 
-- Fig.~\ref{qcd}b and   Fig.~\ref{qcd}d, respectively.
The correction sizes   are  at the level of few percent. 
If the partonic  $k_T$ effects are neglected,
the correction functions are flat. Their inclusion modifies 
sizably the  $C_{QCD}(\rho_l)$ function
which is highly sensitive to the relative shape of the partonic  
$k_T$ distribution in the $W$ and $Z$-boson  production events.
It is important to note,  that the above corrections must be  determined and 
applied at  the raw-data level. 
This is an important merit of the 
proposed correction method which, by construction, 
takes care of the QCD effects independently 
of the level of understanding of the detector performance.

 In Fig.~\ref{qcd1}  we show the uncorrected ratios $R_{WZ}^{iso}$ 
for the $\eta_l$ and $\rho_l$ distributions 
-- Fig.~\ref{qcd1}a and Fig.~\ref{qcd1}c, respectively, and the ratio 
corrected for the relative QCD effects in $W$ and $Z$-boson production  
\begin{equation}
 R_{WZ} = R_{WZ}^{iso} \; C_{QCD}\,,
\label{eq:RWZ}
\end{equation}
 Fig.~\ref{qcd1}b and   Fig.~\ref{qcd1}d, respectively, for the
$\eta_l$ and $\rho_l$ distributions.
  We propose the latter ratio to  be used for high precision determination of  
  the  $M_W$ and $\Gamma_W$ parameters at LHC. The estimation of the 
  achievable precision of such a method will be discussed in the separate  
   paper~\cite{Wmass}.
 
It remains to be  noted that the QCD-effects corrected ratios are insensitive 
to the precision of relative normalisation of the two data sets taken 
at the energies $\sqrt{s_1}$ and $\sqrt{s_2}$. Their final precision will 
 be limited entirely  by the statistical accuracy of the lepton-pair event 
samples.
The basic merit of the observables introduced in Section~\ref{reduction} 
is that their QCD-correction factors
are sufficiently small to make the iteration of the electroweak parameters, 
discussed earlier
in this section,  unnecessary for the measurement precision down 
to the level of ${\cal O}(10^{-4})$.

\section{Conclusions }
\label{conclusions}

In this  paper we  have proposed a  strategy of using  the $Z$-boson 
as {\em ``the  standard candle''} for the high-precision measurements of the 
$W$-boson observables  at the LHC. Our goal was 
to propose  the measurement method  and to define the dedicated 
observables which are insensitive to  the ambiguities in  
modelling of the colour and flavour degrees of freedom
of the effective partonic beams.
In addition,  our goal was to minimise the impact 
of those of systematic errors that affect  
differently the $W$ and $Z$-boson production processes.
We have demonstrated that the effect of the uncertainties  
in the partonic distribution functions
can be reduced from the 
level of $5\%$ for the standard observables to  the level of
 $2 \times 10^{-4}$ 
for the observables proposed  in this paper, while preserving 
their sensitivity to the SM parameters. We have demonstrated
that the sensitivity of the proposed observables to the scale error 
of the reconstructed lepton 
momentum, the dominant source of systematic error of the measured
$W$-boson mass, can be reduced  by the factor of $10$.
We have defined the measurement 
procedure in which the relative effects of strong interaction of quarks 
producing the
$W$ and $Z$-bosons are factorised out and measured directly.
Such a  procedure could  allow us to determine the SM
parameters with no need of  modelling the 
perturbative and non-perturbative QCD effects in the Monte Carlo generators.
 The methods developed in this paper will be used 
 in the dedicated studies of the achievable precision of the SM parameters 
measurements at LHC. 
These studies will be reported in the separate papers.

\vspace{8mm}
\noindent
{\large\bf Acknowledgements}
\vspace{3mm}

\noindent
We would like to thank G.\ Calderini and M.\ Seymour for the careful reading
of the manuscript and useful remarks.

\vspace{5mm}
\noindent

\bibliographystyle{h-physrev3.bst}
\bibliography{Refs}

\end{document}